\documentclass[prd,aps,preprint,nofootinbib,superscriptaddress]{revtex4}
\usepackage{epsfig}
\usepackage{amsmath}
\input{epsf}

\begin{document}


\vspace*{2cm}
\title{A note on the scale dependence of the Burkardt sum rule}

\author{Jian Zhou}
\affiliation{\normalsize\it Nikhef and Department of Physics and Astronomy, VU University Amsterdam,  De Boelelaan 1081, NL-1081 HV Amsterdam, The Netherlands}

\begin{abstract}
In this short note, we argue that the Burkardt sum rule for the Sivers functions
can be used to check the consistency of evolution equations of three-parton correlators.
\end{abstract}

\maketitle

Sum rules are often very useful for providing model independent constraints for non-perturbative objects.
Among them, the Burkardt sum rule~\cite{Burkardt:2003yg} is of particular interest in studying transverse single spin asymmetry(SSA) phenomenology.
It states that the net transverse momentum carried by partons inside a transversely polarized nucleon
vanishes when summing over all parton flavors,
\begin{eqnarray}
\sum_{a=q, \bar q, g} \langle k_\perp^a \rangle =0
\end{eqnarray}
Expressed in terms of the Sivers functions~\cite{Sivers:1989cc} it takes form,
\begin{eqnarray}
\sum_{a=q, \bar q, g} \int_0^1 dx \int d^2  k_\perp \  k_\perp^2 f_{1T}^{\perp a}(x,k_\perp^2)=0
\end{eqnarray}
The Burkardt sum rule was first derived in Ref.~\cite{Burkardt:2003yg},
 and previously has been proved in an alternative way by Lorce~\cite{Lorce:2015lna}.
 It has been checked in various
 model calculations~\cite{Goeke:2006ef,Courtoy:2008vi,Courtoy:2008dn,Pasquini:2010af,Pasquini:2011tk}.

Using the well known tree level relations between the $ k_\perp^2$ moment of the Sivers functions
and the corresponding collinear twist-3 correlations~\cite{Boer:2003cm}, the sum rule can be  re-expressed as,
\begin{eqnarray}
\int_0^1 dx T_G^{(+)}(x,x)+\sum_{a=q, \bar q} \int_0^1 dx T_F^a(x,x)=0
\end{eqnarray}
where $T_F$ is the Qiu-Sterman(QS) function~\cite{Efremov:1981sh,Qiu:1991pp},
and $T_G^{(+)}$ is the C-even tri-gluon correlation~\cite{Ji:1992eu,Kang:2008qh,Beppu:2010qn}.
The scale dependence of the sum rule thus can be investigated by using the scale evolution equations of the relevant twist-3
 correlations $T_F$ and $T_G^{(+)}$ which already existed in the  literatures
~\cite{Kang:2008ey,Zhou:2008mz,Vogelsang:2009pj,Braun:2009mi,Ma:2011nd,Schafer:2012ra,Ma:2012xn,Kang:2012em,Sun:2013hua,Schafer:2013wca,Schafer:2013opa,Dai:2014ala}.
 Note that this subject has also been studied from a different aspect of view in Ref.~\cite{Ratcliffe:2014nla}.

We start from the scale evolution equation for quark QS function
$T_F^q(x,x)$~\cite{Kang:2008ey,Zhou:2008mz,Vogelsang:2009pj,Braun:2009mi,Ma:2011nd,Schafer:2012ra,Ma:2012xn,Kang:2012em,Sun:2013hua},
\begin{eqnarray}
&&\!\!\!\!\!\!\!\!\!\!\!\!\!\!\!\!\!
\frac{\partial T_F^q(\xi,\xi,\mu^2)}{\partial {\ln\mu^2}}|_{q, \bar q\rightarrow q}
=\frac{\alpha_s}{2\pi} \int_{\xi} \frac{dx}{x} \left [
C_F \left \{ \frac{1+z^2}{(1-z)_+}+ \frac{3}{2} \delta(1-z) \right
\} T_F^q(x,x) \right .\
\nonumber \\
&& + \left .\ \!\!\! \frac{C_A}{2} \left \{
\frac{1+z}{1-z } T_F^q(\xi,x) -\frac{1+z^2}{1-z}  T_F^q(x,x)-2\delta(1-z)T_F^q(x,x) \right \} \right .\
\nonumber \\
&&- \left .\ \!\!\!
\frac{N_c}{2} \tilde T_F^q(\xi,x)+\frac{1}{2N_c} (1-2z) T_F^q(\xi, \xi-x)
-\frac{1}{2N_c}\tilde T_F^q(\xi,\xi-x)
\right ]
\end{eqnarray}
where $z=\xi/x$. The last two terms in the above formula contain anti-quark contribution.
We proceed by carrying out the integration over $\xi$ on both sides of the above equation,
\begin{eqnarray}
\frac{\partial T_F^q(\mu^2)}{\partial {\ln\mu^2}}|_{q, \bar q \rightarrow q}
&=& \!\!
\frac{\alpha_s}{2\pi}   \frac{C_A}{2} \left \{
\int dx  dz \frac{1+z}{1-z } T_F^q(zx,x) -T_F^q(\mu^2)\int dz \frac{1+z^2}{1-z}
\right .\
\nonumber \\ &&
\left .\ \ \ \ \ \ \ \ \ \ \ \
-\int dx  dz \tilde T_F^q(zx,x)
-2T_F^q(\mu^2) \right \}
\nonumber \\ &+& \!\!
\frac{\alpha_s}{2\pi}\frac{1}{2N_c} \int dx  dz \left \{ (1-2z) T_F^q(zx,(z-1)x)
- \tilde T_F^q(zx,(z-1)x) \right \}
\end{eqnarray}
where we introduce the short hand notation $T_F^q(\mu^2)=\int d\xi \ T_F^q(\xi,\xi, \mu^2)$.
The same analysis applies to the scale evolution of anti-quark QS function $T_F^{\bar q}(\mu^2)$.
Note that our convention for $T_F^{\bar q}$ differs from the one used in Ref.~\cite{Kang:2008ey} by a minus sign,
 such that the relation between the $k_\perp^2$ moment of the anti-quark Sivers function and $T_F^{\bar q}$ is the same as the
 relation for the quark ones in a same process(for example, SIDIS process).
Combining quark and anti-quark contributions, one obtains,
\begin{eqnarray}
\frac{\partial \left [ T_F^q(\mu^2)+ T_F^{\bar q}(\mu^2)\right ]}{\partial {\ln\mu^2}}|_{q,\bar q \rightarrow q,\bar q}
&=&
\frac{\alpha_s}{2\pi}   \frac{C_A}{2} \left \{
\int dx  dz \frac{1+z}{1-z } \left [ T_F^q(zx,x)+T_F^{\bar q}(zx,x)] \right ]
\right .\
\nonumber \\ &&    \left .\
 -\int dz \frac{1+z^2}{1-z} \left [ T_F^q(\mu^2)+T_F^{\bar q}(\mu^2) \right ]
 \right .\
 \nonumber \\ & & \left .\ \!\!\!\!\!\!\!\!\!\!\!\!\!\!\!\!\!\!\!\! -
\int dx  dz \left [\tilde T_F^q(zx,x)+\tilde T_F^{\bar q}(zx,x) \right ]-
 2\left [ T_F^q(\mu^2)+T_F^{\bar q}(\mu^2) \right ] \right \}
\label{qq}
\end{eqnarray}
To arrive the above result, we have made use of the symmetry properties $T_F^{q,\bar q}(x_1,x_2)=T_F^{q,\bar q}(x_2,x_1)$,
 $\tilde T_F^{q,\bar q}(x_1,x_2)=-\tilde T_F^{q,\bar q}(x_2,x_1)$ and the relations $T_F^q(x_1,-x_2)=T_F^{\bar q}(-x_1,x_2)$,
 $\tilde T_F^q(x_1,-x_2)=\tilde T_F^{\bar q} (-x_1,x_2)$~\cite{Kang:2008ey,Braun:2009mi}.

The scale evolution of the tri-gluon correlation $T_G^{(+)}$ also receives the contribution
from quark and anti-quark~\cite{Braun:2009mi,Schafer:2013wca}
\begin{eqnarray}
\frac{\partial T_G^{(+)}(\xi,\xi,\mu^2)}{\partial {\ln\mu^2}}|_{q, \bar q\rightarrow g}
&=&\frac{\alpha_s}{2\pi} \sum_{q ,\bar q} \int_{\xi} \frac{dx}{x} \frac{C_A}{2} \left \{
\frac{1+(1-z)^2}{z } \left [ T_F^q(x,x) +T_F^{\bar q}(x,x)\right ] \right .\
\nonumber \\ && \ \ \ \ \ \ \ \ \ \ \ \ \ \ \ \ \ \ \ \  \left .\
-\frac{2-z}{z}  \left [ T_F^q(x,x-\xi)+T_F^{\bar q}(x,x-\xi) \right ] \right .\
\nonumber \\ &&  \ \ \ \ \ \ \ \ \ \ \ \ \ \ \ \ \ \ \ \ \left .\
+\left [ \tilde T_F^q(x-\xi,x)+\tilde T_F^{\bar q}(x-\xi,x) \right ]
\right \}
\end{eqnarray}
Carrying out the integration over $\xi$ on both sides of the equation and changing the
integration variable $z \rightarrow 1-y$,
\begin{eqnarray}
\frac{\partial T_G^{(+)}(\mu^2)}{\partial {\ln\mu^2}}|_{q, \bar q\rightarrow g}
&=&\frac{\alpha_s}{2\pi}\sum_{q ,\bar q} \frac{C_A}{2}
\left \{ \left [ T_F^q(\mu^2) +T_F^{\bar q}(\mu^2)\right ]
\int dy \frac{1+y^2}{1-y }  \right .\
\nonumber \\ && \ \ \ \ \ \ \ \ \ \
\left .\
+ \int dx \int dy \left (
-\frac{1+y}{1-y}  \right )  \left [ T_F^q(x,yx)+T_F^{\bar q}(x,yx) \right ] \right .\
\nonumber \\ && \ \ \ \ \ \ \ \ \ \
\left .\
+ \int dx \int dy
\left [ \tilde T_F^q(yx,x)+\tilde T_F^{\bar q}(yx,x) \right ]
\right \}
\label{qg}
\end{eqnarray}
Adding up Eq.~\ref{qq} and Eq.~\ref{qg}, we obtain,
\begin{eqnarray}
\frac{\partial \sum_{q ,\bar q} \left [ T_F^q(\mu^2)+ T_F^{\bar q}(\mu^2)+T_G^{(+)}(\mu^2) \right ] }{\partial {\ln\mu^2}}|_{q, \bar q \rightarrow q, \bar q,g}
=- \frac{\alpha_s}{2\pi} \sum_{q ,\bar q} C_A \left [ T_F^q(\mu^2)+ T_F^{\bar q}(\mu^2) \right ]
\label{qqg}
\end{eqnarray}

We now consider the contribution from the tri-gluon correlation
$T_G^{(+)}(x,x)$ to the scale evolution of the  quark  and anti-quark QS functions.
It is worthy to mention that
the O type tri-gluon correlation(related to the function $T_G^{(-)}(x,x)$)  contributes the same in magnitude
and opposite in sign to both the quark and anti-quark Sivers functions.
This is not  surprising  because the O type tri-gluon relation is related to the
$k_\perp$ moment of the spin dependent odderon which measures the difference between particle and anti-particle scattering~\cite{Zhou:2013gsa}.
Therefore, to study the scale dependence of the sum rule, we only need to consider $T_G^{(+)}(x,x)$
contribution which has been given in Refs.~\cite{Kang:2008ey,Ma:2012xn,Dai:2014ala},
\begin{eqnarray}
\frac{\partial T_F^q(\xi,\xi,\mu^2) }{\partial {\rm ln} \mu_F^2} |_{g \rightarrow q}
   &=& \frac{\alpha_s}{2\pi}  \int_\xi \frac{d x}{x} \frac{1}{2} \left[ z^2+(1-z)^2 \right ]
T_G^{(+)}(x,x)
\end{eqnarray}
which leads to
\begin{eqnarray}
\frac{\partial T_F^q(\mu^2) }{\partial {\rm ln} \mu_F^2} |_{g \rightarrow q}
   &=& \frac{\alpha_s}{2\pi} \frac{1}{3}
T_G^{(+)}(\mu^2)
\label{gq}
\end{eqnarray}
The same equation applies to the anti-quark case.

We move on to discuss the contribution from the tri-gluon correlation $T_G^{(+)}$ to the  scale evolution of itself.
Three different evolution equations have been derived by three groups, respectively~\cite{Kang:2008ey,Braun:2009mi,Schafer:2013opa}.
The possible source of the discrepancy between the results in Refs.~\cite{Kang:2008ey,Schafer:2013opa} is that the different parametrization for
the tri-gluon correlations have been used~\cite{Kang:2008qh,Beppu:2010qn}. We are not able to localize the source of the discrepancy between the results~\cite{Braun:2009mi,Schafer:2013opa},
as the authors of paper~\cite{Braun:2009mi} derived the evolution equations in a very different formalism.
Here, we use the one obtained in Ref.~\cite{Schafer:2013opa},
\begin{eqnarray}
\frac{\partial \left [ \frac{N(\xi,\xi)- N(\xi,0)}{\xi} \right ]}{\partial {\rm ln} \mu_F^2} |_{g \rightarrow g}
   &=&
\frac{\alpha_s}{2\pi}C_A  \int_\xi^1 \frac{d x}{x^2}
\left \{
 \frac{(z^2-z+1)^2}{z(1-z)_+} \left [ N(x,x)
 -  N(x,0)\right ] \right .\
\nonumber \\ &&
 + \frac{1+z^2}{2z(1-z)_+} N(\xi,x)
  - \frac{1+(1-z)^2}{2z(1-z)_+} N(x,x-\xi)
    \nonumber \\ && \left .\ \!\!\!
 -\frac{z^2+(1-z)^2}{2z(1-z)_+}N(\xi,\xi-x)
 - \delta(1-z)\left [N(x,x)-N(x,0)\right ] \right \}
 \nonumber \\ && + \frac{\alpha_s}{2\pi}
 \left ( C_A \frac{11}{6}-\frac{n_f}{3} \right ) \left [N(\xi,\xi)-N(\xi,0)\right ]
 \label{gequation}
\end{eqnarray}
where the combination of the N type tri-gluon correlation defined in Ref.~\cite{Beppu:2010qn} can be related to
$T_G^{(+)}$ through the following identity,
\begin{eqnarray}
 \frac{x}{2 \pi} T_G^{(+)}(x,x)&=&-4M_N \left [ N(x,x)- N(x,0) \right ]
\end{eqnarray}
Carrying out the integration over $\xi$ on the both sides of Eq.~\ref{gequation}, we obtain,
\begin{eqnarray}
\frac{T_G^{(+)}(\mu^2)}{\partial {\rm ln} \mu_F^2} |_{g \rightarrow g}
   =
-\frac{\alpha_s}{2\pi}C_A T_G^{(+)}(\mu^2)
-\frac{\alpha_s}{2\pi} \frac{n_f}{3}  T_G^{(+)}(\mu^2)
 \label{gg}
\end{eqnarray}
To arrive the above compact expression, we have used the same mathematic trick, i.e. changing integration variable
$z\rightarrow 1-y$, as well as the symmetry properties for the N type tri-gluon correlation:
$ N(x_1,x_2) = N(x_2, x_1) , \ \ N(x_1,x_2) = -N(-x_1,-x_2)$~\cite{Beppu:2010qn}.

Collecting the intermediate results presented in Eq.~\ref{qqg}, Eq.~\ref{gq}, and Eq.~\ref{gg}, one obtains,
\begin{eqnarray}
\frac{\partial \sum_{q,\bar q,g} \left [ T_F^q(\mu^2)+ T_F^{\bar q}(\mu^2)+T_G^{(+)}(\mu^2) \right ] }{\partial {\ln\mu^2}}
=- \frac{\alpha_s}{2\pi}  C_A \sum_{q,\bar q,g} \left [ T_F^q(\mu^2)+ T_F^{\bar q}(\mu^2) +T_G^{(+)}(\mu^2)\right ]
\end{eqnarray}
It is worthy to mention that all terms on the right side of the equation entirely
come from the so-called boundary terms first discovered by Braun, Manashov and Pirnay~\cite{Braun:2009mi}.
We reexpress the above formula as,
\begin{eqnarray}
\frac{\partial  \left [ \sum_{a=q, \bar q, g} \langle k_\perp^a \rangle (\mu^2) \right ] }{\partial {\ln\mu^2}}
 =- \frac{\alpha_s}{2\pi} C_A \sum_{a=q, \bar q, g} \langle k_\perp^a \rangle(\mu^2)
\end{eqnarray}
which can be readily solved,
\begin{eqnarray}
\sum_{a=q, \bar q, g} \langle k_\perp^a \rangle (Q^2)
= e^{-\frac{C_A}{2\pi}\int_{Q_0^2}^{Q^2} \alpha_s(\mu^2) \frac{d \mu^2}{ \mu^2 }}
\sum_{a=q, \bar q, g} \langle k_\perp^a \rangle (Q^2_0)
\label{final}
\end{eqnarray}
This is the main result of our short note.
It is easy to see that the Burkardt sum rule holds at any scale provided that it is satisfied  at one
arbitrary scale. If the Burkardt sum rule were violated at a low initial  scale for some unknown reason, judging from the
above formula, it will be asymptotically  satisfied at higher scale.

To summarize: We found that the Burkardt sum rule is stable under QCD
scale evolution if one uses the  tri-gluon evolution equation obtained in Ref.~\cite{Schafer:2013opa}.
However, without making any model dependent assumption for the function $T_G^{(+)}$, we are not able to show that the Burkardt sum rule is  preserved under  QCD scale evolution when
the different tri-gluon evolution equations are used~\cite{Kang:2008ey,Braun:2009mi}.
These findings might shed new light on the controversy  about the tri-gluon scale evolution.

\

\noindent
{\bf Acknowledgments:}
I am grateful to A. Metz for initiating this work and the critical reading of the manuscript.
I also would like to thank D. Boer and P. Mulders for helpful discussions and valuable  comments.
This research has been supported by the EU "Ideas" program QWORK (contract 320389).

\end {document}